\documentclass[useAMS,usenatbib,a4]{mn2e}
\usepackage[dv ips]{graphicx}

\title[Stochasticity, a variable upper-mass limit, binaries and
  SFRIs]{Stochasticity, a variable stellar upper-mass limit, binaries
  and star-formation rate indicators} \author[John J. Eldridge]{John
  J. Eldridge$^{1,2}$ \thanks{E-mail: jje@ast.cam.ac.uk}
  \\ $^{1}$Department of Physics, University of Auckland, Private Bag
  92019, Auckland, New Zealand\\ $^{2}$Institute of Astronomy, The
  Observatories, University of Cambridge, Madingley Road, Cambridge,
  CB3 0HA\\}

\pagerange{\pageref{firstpage}--\pageref{lastpage}} \pubyear{2005}
\begin{document}
\maketitle
\label{firstpage}

\begin{abstract}
Using our Binary Population And Spectral Synthesis (\textsc{BPASS})
code we explore the effects on star-formation rate indicators of
stochastically sampling the stellar initial mass function, adding a
cluster mass dependent stellar upper-mass limit and including binary
stars. We create synthetic spectra of young clusters and star-forming
galaxies and compare these to observations of H$\alpha$ emission from
isolated clusters and the relation between H$\alpha$ and FUV emission
from nearby galaxies. We find that observations of clusters tend to
favour a purely stochastic sampling of the initial mass function for
clusters less than $100M_{\odot}$, rather than the maximum stellar
mass being dependant on the total cluster mass. It is more difficult
to determine whether the same is true for more massive clusters. We
also find that binary stars blur some of the observational
differences that occur when a cluster-mass dependent stellar
upper-mass limit is imposed when filling the IMF. The effect is
greatest when modelling the observed H$\alpha$ and FUV star-formation
rate ratios in galaxies. This is because mass transfer and merging of
stars owing to binary evolution creates more massive stars and stars
that have greater mass than the initial maximum imposed on the stellar
population.
\end{abstract}


\section{Introduction}

A key problem when modelling stellar populations is how to determine
the distribution of initial stellar masses in the population. The
conventional method is to define an initial mass function (IMF)
according to which the number of stars of a given mass is calculated
as a function of the mass. Typically a power-law of the mass is
used. The first was suggested by \citet{salpeter} where $dN(M) \propto
M^{-2.35} dM$ for $0.3<M/M_{\odot}<10$. Despite being 57 years old this
IMF is still widely used and appears to be universal. This slope holds
over a wide range of stellar masses, only flattening in gradient below
stellar masses of around $1M_{\odot}$
\citep{oldimf,kroupa1,chabrier,bastian}.

The IMF provides the distribution of stellar initial masses in a
stellar population such as that found in stellar clusters. However
when trying to simulate the stellar population in a galaxy it is
important to recognise that a galaxy is not made up of one unique
stellar population. A galaxy is actually made up of numbers of stellar
clusters each with their own mass and age. The masses of these
clusters are also described by their own cluster initial mass
function.  Furthermore fluctuations in the IMF of these clusters,
especially if their mass is less than $10^4 M_{\odot}$, can provide
large variation in the ionising fluxes from the cluster
\citep{cervino1,cervino2,cervino3}. The implication of stars forming
in clusters is that the stellar IMF (SIMF) does not apply across an
entire galaxy. Instead to get the galaxy-wide distribution of stellar
masses we must model a number of stellar clusters with different
masses according to a cluster initial mass function (CIMF) and within
each clusters apply a SIMF to produce an integrated galaxial initial
mass function (IGIMF) \citep[e.g.][]{weidner1,pflamm}. This
is the case even if not all the clusters are dense enough to remain
bound over their lifetimes \citep{pz,bressert,gieles}. \citet{bastian}
and \citet{haas} reviewed and investigated the importance of combining
a CIMF and a SIMF to make a galaxy-wide SIMF. They found that the
resultant galaxy-wide SIMF is most sensitive to the minimum mass for a
cluster, the slope of the CIMF and whether the mass of the most
massive star is limited by the cluster mass.

The most extreme articulation of the last factor is whether it is
physically possible for a $100M_{\odot}$ cluster to be composed of a
single $100M_{\odot}$ star or the more probable scenario of a cluster
composed of many lower-mass stars.. The alternative is that such
low-mass clusters can only put say 10 percent or less of their total
mass into the most massive star. If the former is possible it would
suggest that some O stars might form in isolation (here taken to be in
isolation with no other O or B stars in the same cluster so the other
cluster members are of a much lower mass) and this would have
important implications for the star-formation process. I.e. is star
formation a pure-stochastic process or a bottom-up process with
low-mass stars formed first and high-mass stars only formed if there
is enough material left. After accounting for runaway O stars
contaminating the apparent number of isolated (with no companions at
all) O stars \citet{dewit} suggested that at most $4\pm2$ per cent of
O stars form outside a cluster environment. \citet{parker} considered
that an isolated O star only meant there was no other OB star in the
same cluster. This is the case when a $100M_{\odot}$ cluster is
composed of one star that contains most of the mass of the cluster
with a few very low-mass companions. They modelled the populations of
star clusters using a standard CIMF with a slope of $-2$ and predicted
that 5 per cent of O stars formed in clusters that had no other O or B
stars and would be observed as isolated when in fact they are just
massive stars that have been able to form in a low-mass cluster.

The argument against the conclusion that O stars can form in low-mass
clusters has been presented by \citet{vanbev}, \citet{weidner1}
and \citet{weidner3} who have suggested that observations indicate
that, for a specific cluster mass, there is a maximum possible stellar
mass well below the total cluster mass. If this were the case then,
from their model, a $100M_{\odot}$ cluster could not form stars with
masses above $10M_{\odot}$. Therefore when a synthetic galaxy is
created from the convolution of cluster and stellar IMFs there would
be a dearth of the most massive stars compared with when there are no
restrictions on the maximum stellar mass. \citet{pflamm,pflamm2009}
have suggested that there are differences in star-formation rate
indicators when stellar populations are modelled by the two different
IMF filling methods. However the observations used consider the
differences that occur at low star-formation rates and thus
uncertainties and low-number fluctuations between observed systems
make it difficult to determine whether the maximum stellar mass
depends on the cluster mass. However the complementary studies by
\citet{elmegreen}, \citet{parker} and \citet{masch} show that similar
observations indicate that there is no evidence for restrictions on
the maximum stellar mass in clusters.

In this paper we examine recent observations \citep{lee,lamb,calzetti}
that may provide a firmer constraint on how nature selects the masses
of stars in clusters and galaxies
\citep{pflamm,pflamm2009,cervino3,slug,danw}. The novel feature of
this work is that we are able to demonstrate how binary stars alter
our population synthesis predictions. Recent observations
\citep{bin1,bin2,bin3} indicate that the binary fraction in young
massive stellar populations is close to one. It is therefore vital to
include binary stars especially those that interact. In our
populations approximately two thirds of binaries interact. We first
outline our stellar evolution models and the method of our spectral
synthesis. We then describe our two ways to determine the distribution
of initial masses in synthetic clusters and galaxies. Next we discuss
the observational implications of varying the IMF-filling method on
the H$\alpha$ and FUV star-formation rate indicators. Finally we
present our conclusions.

\section{Numerical Method}

\subsection{Binary population and spectral synthesis}

We have developed a novel and unique code to produce synthetic stellar
populations that include binary stars \citep{EIT,es09}. While similar
codes exist our Binary Population and Spectral Synthesis (BPASS) code
has three important features each of which set it apart from other
codes and enable it to study stochastic effects on the IMF. First, and
most important, is the inclusion of binary evolution when modelling
the stellar populations. The general effect of binaries is to cause a
population of stars to look bluer at older ages than predicted by
single-star models. Secondly, a large number of detailed stellar
evolution models are used to create the synthetic populations rather
than an approximate rapid population synthesis method. Thirdly, we use
as many theoretical inputs in our synthesis with as few empirical
inputs as possible to create a completely synthetic model to compare
with observations.

BPASS uses approximately 15,000 detailed stellar models calculated by
the Cambridge \textsc{STARS} code as described by \citet{EIT}. These
include single star and binary models with initial masses between
$0.5$ and $120M_{\odot}$ and $5$ and $120M_{\odot}$ respectively. We
take $120M_{\odot}$ to be our most massive star possible because of
our limited grid of binary evolution models. Above this mass the
mass-loss rates at solar metallicity on the main-sequence are high
\citep{vinkmdot} and the evolutionary timescales of the stars vary
little as the initial mass is increased further. We note that, owing
to stars merging our binary populations include some single stars that
have effective initial masses of $200M_{\odot}$ and above. The minimum
binary primary mass of $5M_{\odot}$ is selected because initially our
binary models were specifically created to study the progenitors of
core-collapse supernovae. The main-sequence lifetime of a $5M_{\odot}$
star is $100$Myrs, which is the period we use for the duration of the
star-burst in our constant star-formation models so there is no effect
from low-mass binaries in these models. Furthermore observations
indicate the binary fraction decreases at the low masses
\citep{binf1,binf2,binf3}. However the binary fraction is more
complicated than we assume here and is determined by a star cluster's
dynamics, environment and age. It is thought that stars of all masses
can form in binary or multiple systems but these can be broken up by
dynamical interactions in young clusters
\citep[e.g.][]{binrupt,binfevo}. We note that \citet{uvupturn} found
low-mass binary stars can explain the excess UV flux observed in
Elliptical galaxies. Such systems do not contribute strongly until
1Gyr after formation at which time our estimated UV fluxes should
increase only slightly.  Creating a new grid of low-mass detailed
binary models for inclusion in BPASS is unnecessary for this work to
demonstrate the importance of binary stars.

Here we use models at solar metallicity with a metallicity mass
fraction of $Z=0.02$. We include convective overshooting and a
mass-loss prescription that combines the mass-loss rates of
\citet{vink}, \citet{dejager} and \citet{nugis}. The binary evolution
accounts for Roche-lobe overflow, common-envelope evolution, mass
transfer and neutron-star kicks which affect the survival of binary
stars after a supernova. These models are combined with the stellar
atmosphere spectra of \citet{crow}, \citet{wn} and \citet{basel} to
predict the spectra of the stellar populations.

A significant change we make here, compared with our previous work, is
to break from our previous assumption that the SIMF can be described by
a simple Salpeter law over the entire mass range of stars. Our method
requires us to consider that all stars are born in clusters. The mass
of these clusters is described by a CIMF and the mass distribution of
stars within each cluster is described by the SIMF. This is achieved
by first picking a cluster mass and then filling the cluster with
stars from the SIMF. We model multiple clusters together to create
synthetic galaxies with different star-formation histories but with
the same mean \textit{constant} star-formation rate over a long
period of time.

We use two methods of populating the SIMF for our synthetic
clusters. They differ by whether we limit the maximum stellar mass or
not. Our first method is to assume that any star can occur in any
cluster such that, $M_{\rm max} \le M_{\rm cl}$. I.e. the star cannot
be more massive than the cluster it inhabits. This we refer to as pure
stochastic sampling (PSS) of the SIMF. In this SIMF we assume a
Salpeter slope of -2.35 between 0.5 and 120 $M_{\odot}$ and a slope of
-1.3 between 0.1 and 0.5$M_{\odot}$. It is similar to the constrained
sampling method outlined by \citet{weidner1} and used by
\citet{cervino3}.

Our second case has the maximum mass of a star in a cluster
dependent on the total mass of the cluster. We use the relation
calculated by \citet{pflamm} which is given by,
\begin{eqnarray}
\lefteqn{ \log_{10} (M_{\rm max}/M_{\odot})= 2.56 \log_{10} (M_{\rm cl}/M_{\odot})} \nonumber\\
 &&\times \big(3.82^{9.17}+(\log_{10} (M_{\rm cl}/M_{\odot}))^{9.17} \big)^{\frac{-1}{9.17}} - 0.38,\nonumber
\end{eqnarray}
where $M_{\rm max}$ is the maximum stellar mass possible in a cluster
of mass $M_{\rm cl}$. We therefore use $M_{\rm max}$ from this
equation as the maximum mass in our initial mass function up to a
limit of 120$M_{\odot}$ in our synthetic clusters. We refer to this
method as the cluster mass dependent maximum stellar mass (CMDMSM)
method. The resulting clusters are similar to those from the
sorted-sampling method outlined by \citet{weidner1}.

We note that our synthetic populations have some limitations. In
Figures \ref{clustermodels} and \ref{galaxymodels} there are diagonal
and horizontal linear features in the distribution of model
populations. These arise at low cluster-masses and star-formation
rates owing to the limited resolution of the stellar model initial
masses and time bins used in our synthesis. This becomes most
noticeable when there is only one massive star in the stellar
population. One solution to this would be to interpolate between
stellar models but given that stellar evolution is non-linear and
binary evolution is even less predictable we avoid spurious results
from interpolations and select the closest model available.

Also our binary population models are not complete and here we are
only demonstrating the importance of including binary stars. For
example, we do not include binaries with initial primary masses below
$5M_{\odot}$ and as yet we do not consider the emission from X-ray
binaries. This would provide another source of ionising flux that
would also effect the H$\alpha$ and UV flux ratio. The effect would be
more important at low cluster masses and low star-formation rates
where one X-ray binary would dominate the entire ionising flux from
the stellar population \citep{xrb}.

\subsection{Creating synthetic clusters}

We use the PSS and CMDMSM methods to create synthetic stellar
populations in two regimes. In the first we consider individual
stellar clusters with all the stars coeval. We create models of
stellar clusters with both PSS and CMDMSM and investigate how they
affect the H$\alpha$ line flux per $M_{\odot}$ in the cluster. Our
process for creating a synthetic cluster to compare to the
observations of \citet{calzetti} is as follows.

\begin{enumerate}
\item We randomly generate a cluster mass between 10 and $10^6M_{\odot}$
  from the CIMF which has a slope of $-2$ \citep{degrijs,ladalada}.
\item We fill the cluster with stars, the masses of which are picked at
  random from the SIMF with the maximum stellar mass given by PSS or
  CMDMSM.
\item We add stars to the cluster until the total mass is greater than
  our target cluster mass. We then consider whether the final cluster
  mass is closer to the target cluster mass with or without the last
  star added to the cluster. If the mass is closer without the last
  star we remove the last star from the cluster. This is similar to
  the sorted sampling of \citet{weidner1} and makes it less likely
  that a star can be added that is more massive than the target
  cluster mass as in the soft sampling of \citet{elmegreen}.
\item We randomly generate the cluster age between 1 and
  8$\,$Myr. This is to match the observed age range of
  \citet{calzetti}.
\item We calculate the H$\alpha$ flux for the resultant stellar
  population. This is done with theoretical stellar atmospheres
  and stellar models to predict the resultant total spectrum as
  described by \citet{es09}. We calculate the number of ionising
  photons from wavelengths shortward of 912\AA\, and convert this to
  the flux of H$\alpha$ by assuming $10^{11.87}$ ionising photons give
  rise to $1 \,{\rm erg \, s^{-1}}$ of H$\alpha$ flux.
\end{enumerate}

This process is repeated for many different cluster masses so that we
can build up a picture of how H$\alpha$ flux varies with cluster mass
for clusters aged between 1 and 8$\,$Myr. \citet{calzetti} performed
an observational study of such clusters and provide the observed mean
H$\alpha$ flux per $M_{\odot}$ for two different masses of
clusters. We compare our models to these observed populations in
Section \ref{calsec}. In the binary population case we include a
companion for every star that has an initial mass greater than
$5M_{\odot}$. We assign binary parameters at random from a flat
initial mass ratio and flat distribution of the logarithm of the
initial separation using the model closest to the parameters from our
grid of models calculated by \citet{EIT}. We include the mass of the
companion in the total cluster mass.

\subsection{Synthetic galaxies}

Our second set of population models are for synthetic galaxies with an
assumed constant star-formation rate. Rather than fill up the
population of a galaxy according to a galaxy-wide IMF we create the
galaxy from a set of clusters that each have their own individual age
and stellar population. To create a galaxy we first pick a
star-formation rate between $10^{-5}$ to $10\,M_{\odot}{\rm
  yr^{-1}}$. We then create the synthetic galaxy as follows.

\begin{enumerate}
\item We pick a cluster mass at random from a CIMF which has a slope
  of $-2$ between 50 and $10^6 M_{\odot}$.
\item We fill the cluster with a stellar population as described in
  Section 2.2 and aged to between 0 and 100 Myr, chosen at random from
  a uniform distribution.
\item We continue this process until the total mass created in the
  galaxy over 100$\,$Myr gives the required star-formation rate.
\item With this stellar population we calculate the number of ionising
  photons from wavelengths shortward of 912\AA\, and convert this to
  the flux of H$\alpha$ by assuming $10^{11.87}$ ionising photons give
  rise to $1 \,{\rm erg \, s^{-1}}$ of H$\alpha$ flux. We also
  calculate the UV flux density at a wavelength of $1500$\AA.
\item From these H$\alpha$ and UV fluxes we calculate an apparent
  star-formation rate from both and find their ratio. We assume a
  star-formation rate of $1M_{\odot} \, {\rm yr^{-1}}$ produces a
  H$\alpha$ flux of $\log_{10}(F({\rm H}\alpha)/{\rm ergs \,
    s^{-1}})=41.1$ and a UV flux density of $\log_{10}(F(1500{\rm \AA})/{\rm
    ergs \, s^{-1}\,Hz^{-1}})=27.85$ as in \citet{K98}.
\end{enumerate}

We perform these simulations for single and binary populations and for
the PSS and CMDMSM methods of filling the IMF so that the differences
can be compared. Here we use a different range of cluster masses based
on the suggestion of \citet{ladalada} that there is a turn-over in the
mass function of molecular clouds at around 50$\,M_{\odot}$. We also
only consider a period of 100 Myr because this is of the order of a
typical star-formation burst duration \citep{mcquinn}. We also find
that increasing the age beyond 100 Myr has little effect on our
results because it is the typical lifetime of stars that contribute to the
FUV. We note that, when used to create a synthetic galaxy, our CMDMSM
method is based on the IGIMF method of \citet{weidner1}. However we do
not limit the maximum cluster mass in a synthetic galaxy by the total
star-formation rate as they do in their IGIMF method. Recent investigations of
the CIMF suggest that there is no such dependence
\citep{cimf1,cimf2}. In this work we wish to concentrate on whether
the maximum stellar mass depends on the cluster mass. We have
calculated IGIMF models to see the effect of including such a limit
and find our models are in agreement with those of \citet{pflamm} and
\citet{pflamm2009}. Also like \citet{slug} we find that IGIMF
synthetic galaxies cannot reproduce the observed spread of
star-formation rate ratios. This is because restricting the maximum
cluster mass decreases the number of massive stars even more
dramatically than they are in our CMDMSM models.

In Section \ref{obs2} we compare the synthetic populations to the
observed galaxies of \citet{lee}. The novel feature of our approach
is in not forcing clusters to form at the same time but allowing each
to have a different age. This leads to much more scatter in the
predicted observables of our synthetic galaxies. This was also found
by \citet{slug} and \citet{danw}. We have also varied the age range
used for the synthetic galaxies and find that increasing the age has
little effect on our results. Using a younger upper age limit
increases the amount of H$\alpha$ flux relative to the UV flux. This
is because the stars that cause H$\alpha$ emission are more massive
and typically have lifetimes of 10$\,$Myr or less, while stars that
contribute to the FUV continuum span a much greater lifetime range of
up to $100\,$Myr.

\begin{figure*}
\includegraphics[angle=0, width=170mm]{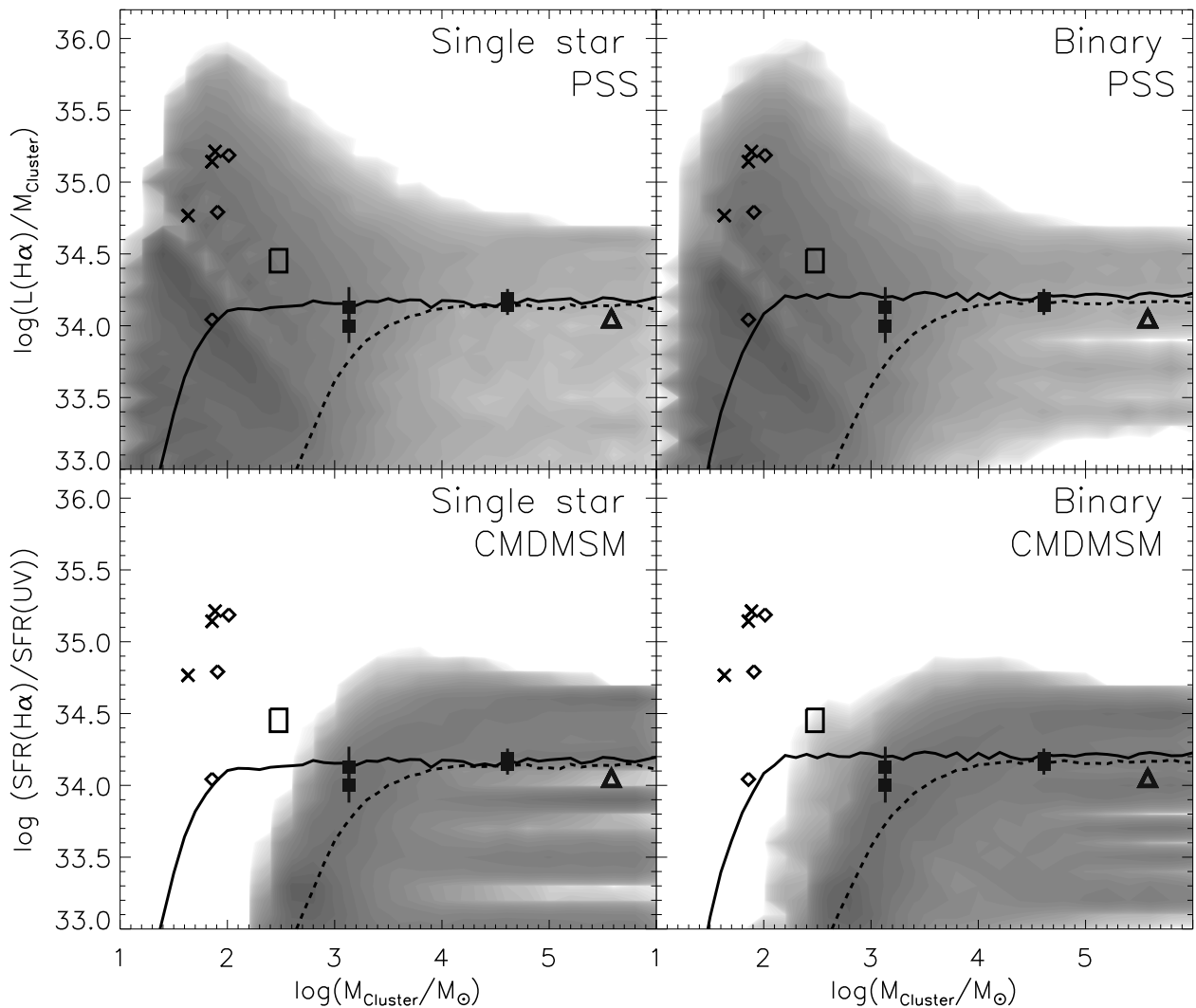}
\caption{H$\alpha$ flux per $M_{\odot}$ for our synthetic
  clusters. The shaded contours represent the density of individual
  realisations of different clusters for two methods of filling the
  IMF: PSS for upper panels and CMDMSM for the lower panels. The lines
  show the mean fluxes from these two methods, the solid black line
  for PSS and dashed black line for CMDMSM. The lines are shown in
  both single star panels and both binary star panels. The black boxes
  represent the observations of \citet{calzetti}, the lower points
  include clusters not detected in H$\alpha$ and the high flux value
  does not include them. The black open box represents the range of
  possible values derived from the Velorum cluster estimated from
  \citet{jefferies} and \citet{demarco}. The diamonds and crosses are
  based on data of \citet{lamb}. The diamonds represent clusters with
  upper limits to their mass and the crosses with measured cluster
  masses. The grey triangle indicates the position of the massive HII
  region NGC604 with values from \citet{eldridgerelano}. The left
  panels are for single-star populations and and the right panels are
  for binary populations.  The linear features in the contours in the
  panels are due to the limited resolution of the stellar models
  initial masses and time bins used in the synthesis.}
\label{clustermodels}
\end{figure*}

\section{Results}

\subsection{The H$\alpha$ from individual clusters}
\label{calsec}

\citet{calzetti} suggested a novel test for determining how the IMF
defines the population of stellar clusters. They studied the
production of ionising photons by young clusters in NGC5194. If an IMF
is populated purely stochastically then one $10^5 M_{\odot}$ cluster
should have the same stellar content as a hundred $1000M_{\odot}$
clusters. Therefore both samples would have the same H$\alpha$ flux
per $M_{\odot}$ of stars. However if the IMF of a $1000M_{\odot}$
cluster is devoid of massive stars due to a link between $M_{\rm max}$
and $M_{\rm Cl}$ then the hundred $1000M_{\odot}$ clusters would have
less H$\alpha$ flux per $M_{\odot}$ than a $10^5 M_{\odot}$ cluster.

For our population of synthetic clusters we have calculated the mean
H$\alpha$ flux per $M_{\odot}$ as for the observed clusters of
\citet{calzetti}. Figure \ref{clustermodels} shows our synthetic
clusters as points along with the mean H$\alpha$ flux. We see that at
cluster masses below $10^4 M_{\odot}$ the results diverge. With PSS it
is possible to have one massive star making up most of the mass of a
cluster while with CMDMSM this is not possible. Therefore for PSS and
CMDMSM the observed mean H$\alpha$ flux drops from the mean value of
around $10^{34.1} {\rm erg \, s^{-1} M_{\odot}^{-1}}$ at around $10^2$
or $10^4 M_{\odot}$ respectively. Therefore by measuring the H$\alpha$
flux for clusters in between these key masses we should be able to
determine how nature fills the IMF.

\citet{calzetti} provide two observed values for their two mass
bins. The first and higher value does not include clusters that are
undetected in H$\alpha$. The second includes these non-detections. The
observations at a cluster mass of $10^{4.5} M_{\odot}$ agree with the
predicted mean $H\alpha$ flux. However the observed points at $10^3
M_{\odot}$ are less conclusive. The point without the non-detections
lies on the the PSS line, while the point including the non-detections
lies in between the PSS and CMDMSM lines. Thus PSS gives a better fit
but a refined CMDMSM scheme that allows a higher maximum mass for a
certain cluster mass may match the \citet{calzetti} data.

An alternative method to discriminate between PSS and CMDMSM is to
search for individual massive stars that are in low-mass clusters. One
example is the Wolf-Rayet star $\gamma$-Velorum, the nearest
Wolf-Rayet star to the Sun in the Galaxy. It is a binary system
containing stars that were initially 35 and 30$M_{\odot}$ in a cluster
with a total mass of between 250 and 350$M_{\odot}$
\citep{demarco,jefferies,gammavel}. We have indicated the location of
this cluster in Figure \ref{clustermodels}. We see that the PSS
clusters overlap with the parameters of this cluster. The CMDMSM
models for a single star population do not reach this region. The
binary CMDMSM models do reach the parameter space for
$\gamma$-Velorum. The small number of such models indicates such
clusters would be rare. This suggests that PSS is more likely to be in
action in nature although a more relaxed form of CMDMSM would also fit
the observed data.

Other more extreme examples of low-mass clusters with a single massive
star were observed by \citet{lamb}. They observed apparently isolated
O stars and found low mass clusters associated with these stars. Using
the stellar and cluster masses derived by \citet{lamb} and estimating
the ionising flux for the massive star we have plotted their clusters
in Figure \ref{clustermodels}. They are only reproduced by our PSS
method.

This agrees with previous studies by \citet{testi1}
\citet{testi2,testi3} and \citet{parker} who use similar
arguments. \citet{masch} also made a detailed study of all available
information and also favour PSS. However \citet{weidner3} performed a
similar analysis and found that for low mass clusters, below
$100M_{\odot}$ PSS is favoured but more massive clusters appears to
have a CMDMSM relation. The observations of \citet{calzetti} do not
currently favour either PSS or CMDMSM. Here we can only agree that PSS
occurs in low-mass, up to $100M_{\odot}$, clusters. For more massive
clusters, of around $1000M_{\odot}$, it is difficult to differentiate
between PSS and CMDMSM. Finally for cluster masses more than
about $10^4M_{\odot}$ the differences are less important.

Finally we note that our results are in line with those of
\citet{cervino3}. They suggest that for cluster masses below $10^{4}
M_{\odot}$ there is a highly asymmetric scatter of the ionising flux
around the mean integrated values from standard synthesis models
because the single most massive star dominates the ionising flux of
the cluster. This manifests itself in our results by the increased
spread in H$\alpha$ flux per $M_{\odot}$ at low cluster masses. We
note that they suggest that PSS is more favoured than CMDMSM.

\begin{figure*}
\includegraphics[angle=0, width=170mm]{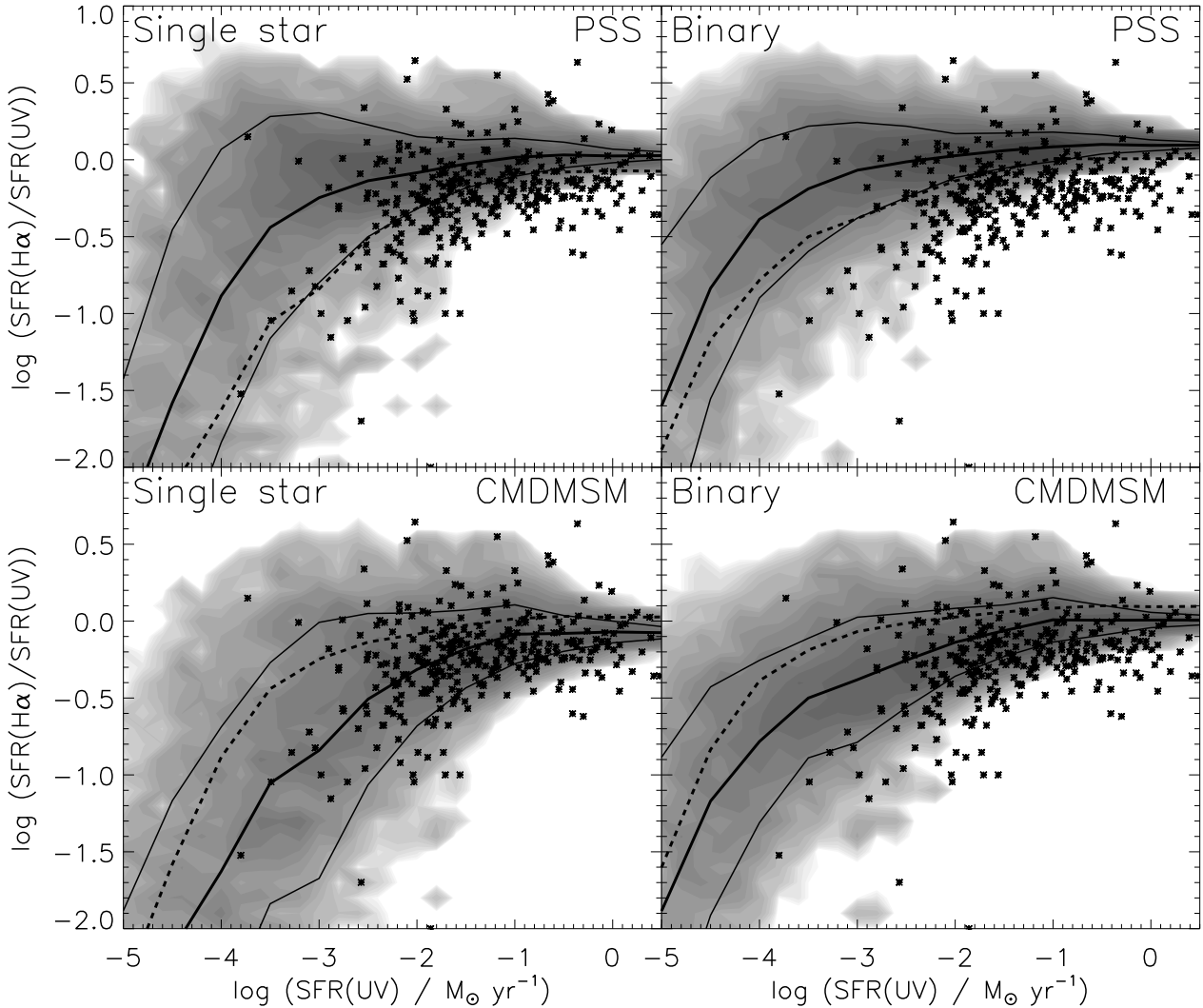}
\caption{The ratio of SFR measured by H$\alpha$ and UV fluxes verses
  the SFR from UV flux. The asterisks are the observations of
  \citet{lee} while the shaded region show the density of our individual
  realisations of synthetic galaxies. The thick solid lines indicate the
  mean ratios for the synthetic galaxies and their $1\sigma$
  limits. The dashed lines show the mean ratios for the other IMF
  filling method with the same stellar population. The upper panels
  are for PSS and the lower panels are for CMDMSM. While the left
  panels are for a single star population and the right panels are for
  binary populations. Here we assume that a star formation rate of
  $1M_{\odot} \, {\rm yr^{-1}}$ is equivalent to a $\log_{10}(F({\rm
    H}\alpha)/{\rm ergs \, s^{-1}})=41.1$ and a UV flux density of
  $\log_{10}(F(1500\AA)/{\rm ergs \, s^{-1}\,\AA})=27.85$. Linear
  features are due to limited resolution in initial mass, separation
  and mass ratio parameter space of our binary models.}
\label{galaxymodels}
\end{figure*}

\subsection{The H$\alpha$ and FUV in 11HUGS galaxies}
\label{obs2}

\begin{figure*}
\includegraphics[angle=0, width=170mm]{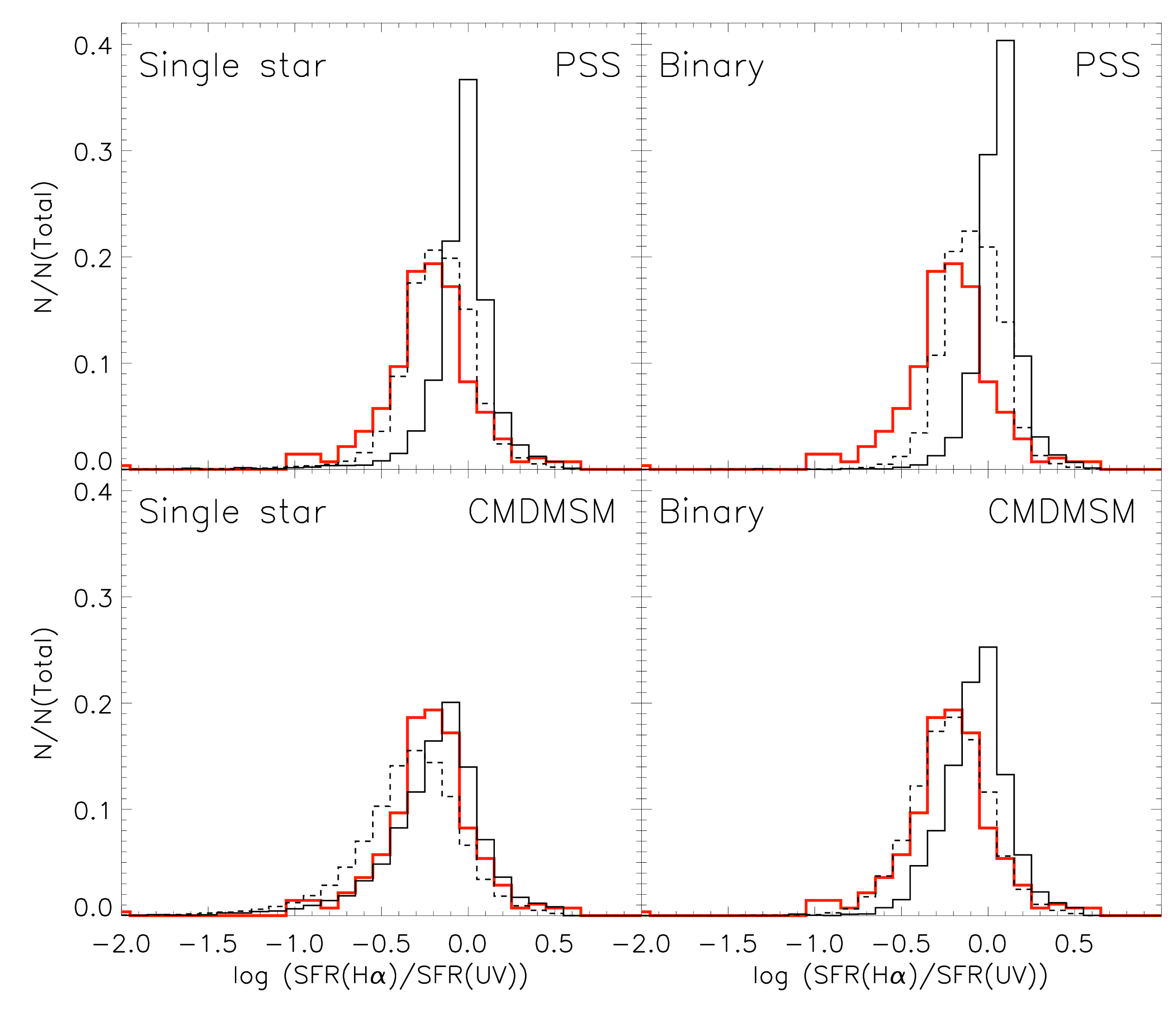}
\caption{The distribution of H$\alpha$ to UV ratio for observed and
  synthetic galaxies with star-formation rates between $10^{-2}$ and 1
  $M_{\odot}{\rm y^{-1}}$. The red line represents the observed sample
  of \citet{lee} while the solid line represents the relevant
  synthetic galaxies from Figure \ref{galaxymodels}. The dashed line
  represents the synthetic observations smeared by a flat leakage of
  ionising photons distributed between a leakage fraction of 0 and 50
  per cent. The left panels are for PSS and the right panels are for
  CMDMSM. The first and third panels are for a single star
  population and the second and fourth panels are for binary
  populations.}
\label{leakage}
\end{figure*}

\begin{figure*}
\includegraphics[angle=0, width=170mm]{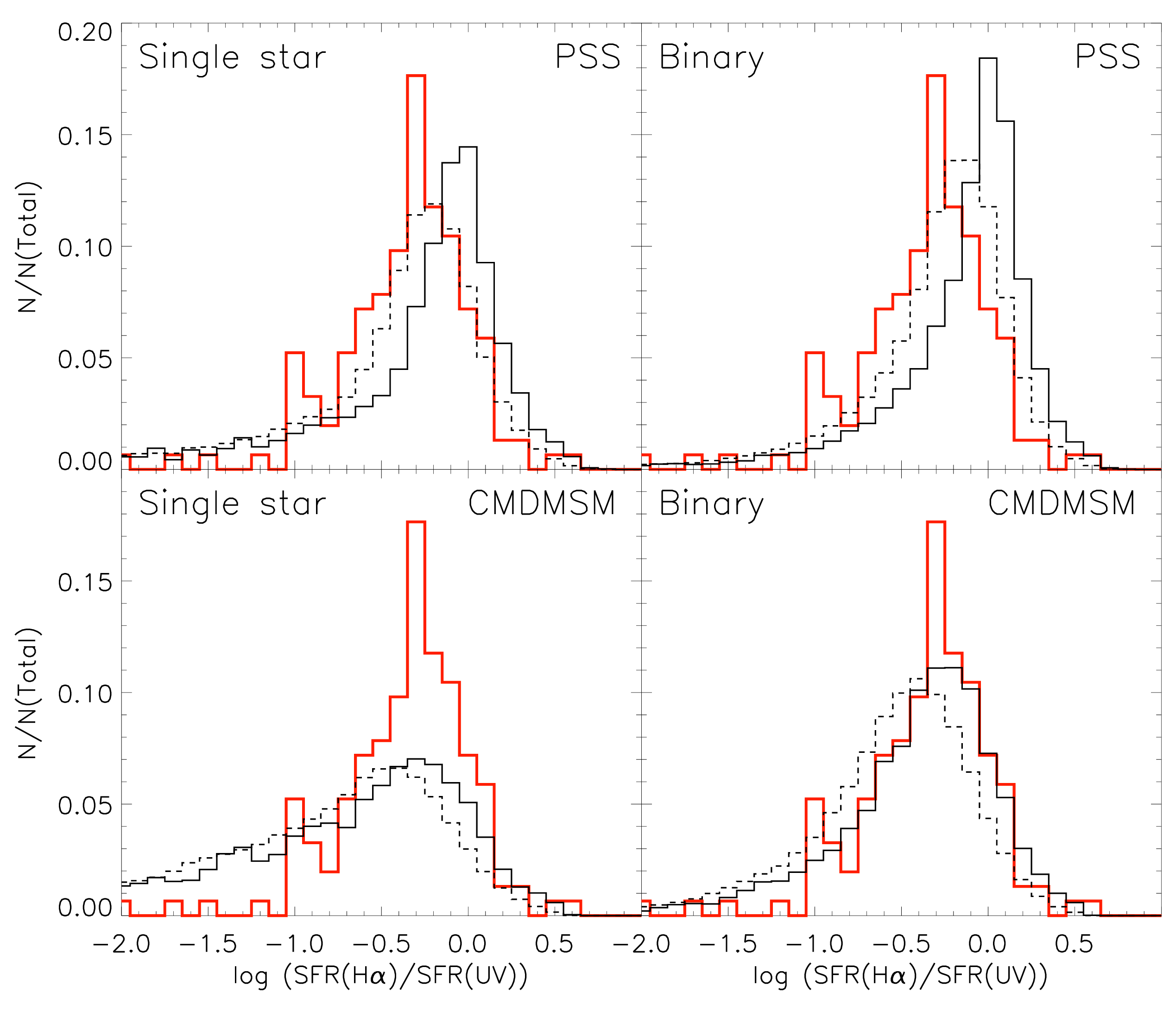}
\caption{The distribution of H$\alpha$ to UV ratio for observed and
  synthetic galaxies with star-formation rates between $10^{-4}$ and
  $10^{-2}$ $M_{\odot}{\rm yr^{-1}}$. The red line represents the
  observed sample of \citet{lee} while the solid line represents the
  relevant synthetic galaxies from Figure \ref{galaxymodels}. The
  dashed line represents the synthetic observations smeared by a flat
  leakage of ionising photons distribution between a leakage fraction
  of 0 and 50 per cent. The left panels are for PSS and the right
  panels are for CMDMSM. While the first and third panels are for a
  single star population and the second and fourth panels are for
  binary populations.}
\label{leakagelow}
\end{figure*}

\citet{lee}, \citet{meurer} and \citet{boselli} have attempted to gain
insight into the IMF by looking at emission from entire galaxies. They
brought together H$\alpha$ observations with far UV continuum
observations. Here we concentrate on the results of \citet{lee}
because their set of galaxies are a volume limited sample of 315
within 11$\,$Mpc. The emission of these two spectral star-formation
rate indicators are determined by the number of stars with masses
greater than 20 and 3$M_{\odot}$ respectively. Therefore measuring the
ratio of the two fluxes, or the relative star-formation rates measured
for the galaxies, gives an indication of the number of stars in
different mass regimes. \citet{lee} found that as the H$\alpha$ flux
decreases the H$\alpha$/UV ratio decreases so there is more UV flux
than expected. \citet{pflamm2009} have suggested that this turn down
is evidence for IGIMF determining the galaxy-wide IMF of these
galaxies.  Their study was based on single star models alone. Here we
repeat their analysis with binary as well as single star models and
also our stochastic approach to the star-formation history, with
stellar clusters forming independently from one another.

We plot our synthetic galaxies in Figure \ref{galaxymodels}. We see
that, at star-formation rates above $10^{-2} M_{\odot}{\rm yr^{-1}}$,
the spread of models is similar but CMDMSM gives a slightly greater
scatter towards lower values of the H$\alpha$/UV ratio. This can be
more easily seen in Figure \ref{leakage} where we bin the synthetic
and observed galaxies with star-formation rates above $10^{-2}
M_{\odot}{\rm yr^{-1}}$ by their H$\alpha$/UV ratio. CMDMSM has a
greater range of ratios because of the relative lack of massive stars
in the total stellar population. We see CMDMSM reproduces the lowest
ratios at the highest star-formation rates. PSS produces much higher
ratios. However in this model we have assumed no leakage of any
ionising photons. This would reduce the contribution from the
H$\alpha$ flux by up to around 50 per cent \citep[see][for
  example]{zurita}. This would lead to lower ratio
values at high star-formation rates for both PSS and CMDMSM.

To account for the leakage or loss of ionising photons from a galaxy
or absorption by dust grains we have made a simple adjustment to our
models. In Figure \ref{leakage} we have modified our synthetic ratio
distributions by assuming that galaxies lose between 0 and 50 per cent
of their ionising photons. We take our synthetic populations and smear
them by this range of possible leakage fraction so that the mean
leakage is 25 per cent. Even for this modest loss of ionising photons
the ratio distribution changes the PSS model to match the range of
observed galaxies. At the same time the CMDMSM method has a slightly
worse agreement. To test the significance of these differences we have
used a $\chi^2$ test to compare the observed distributions to the
synthetic populations. We find that without leakage only CMDMSM with
single stars is a probable match. However with leakage only the CMDMSM
single star synthetic population is ruled out.

\citet{slug} used a leakage fraction of 5 per cent and stated that
their results were not dependent on the amount of leakage.  This is
because they compared the amount of H$\alpha$ to mean values of
H$\alpha$ flux which are less sensitive to leakage than the
H$\alpha$/UV flux ratio (see their figure 2). They found that for
leakage fractions up to 40 per cent their results were
unaffected. This is within the mean leakage of 25 per cent that we
apply to our models.
 
For the single star population the greatest difference between PSS and
CMDMSM is seen in the different paths of the mean ratio values versus
the star-formation rate determined from the UV flux. CMDMSM decreases
much sooner than PSS at around $0.1 M_{\odot}{\rm yr^{-1}}$. However
there is a large possible range around these mean values in both cases
and the difference is only approximately $1\sigma$. When we consider
the binary population we see that the difference between the two IMF
filling methods is substantially reduced. This is because, while the
IMF initially leads to fewer massive stars in the CMDMSM case, binary
interactions, such as merging and mass-transfer, increase the number
of massive stars relative to a single-star population. If we are to
distinguish between PSS and CMDMSM by means of the downturn in this
ratio we must repeat the analysis that led to Figure \ref{leakage} for
lower star-formation rates. We show the result for star-formation
rates between $10^{-2}$ and $10^{-4} M_{\odot}/{\rm yr^{-1}}$ in
Figure \ref{leakagelow}. By eye PSS provides a better fit to the
observed population than CMDMSM. This is because CMDMSM has an
extended tail of galaxies towards lower ratios. PSS does not have this
tail. However including binary stars in the synthetic galaxies reduces
it further in both CMDMSM and PSS. Also the tail might not be present
in the observed data owing to selection biases. A $\chi^2$ test
reveals that both PSS populations are a probable match to the observed
distribution. The single-star CMDMSM distribution does not match the
observed distribution. However our binary CMDMSM population produces
an equally likely fit to the observed data. Our results also show that
some ionising photon leakage is required if our PSS models are to
match observations.

\citet{lee} noted that their results indicate a downturn in the
H$\alpha$ to UV ratio at low star-formation rates. This could be
explained by the IGIMF model put forward by \citet{pflamm2009}. At
first comparison of the synthetic and observed galaxies in Figures
\ref{galaxymodels} and \ref{leakage} tempts us to agree with this
deduction. This is mainly because the spread of the observed galaxies
at higher star-formation rates is better reproduced by the CMDMSM,
single star models. The most significant difference between PSS and
CMDMSM is in the region where the star formation rates drop below
$10^{-2} M_{\odot}{\rm yr^{-1}}$. All our models are able to reproduce
the observed galaxies with the lowest ratios at low star-formation
rates. Therefore it is not possible to differentiate between PSS and
CMDMSM from these observations. Furthermore the inclusion of binaries
in stellar population models means that any difference between PSS and
CMDMSM is only apparent at star-formation rates below those in the
observed sample of \citet{lee}. Therefore, from the observed
distribution of H$\alpha$ to FUV ratio, it is not possible to
discriminate between PSS and CMDMSM owing to the uncertainties in the
importance of binary evolution and ionising photon leakage.

Our conclusions are broadly in line with those of
\citet{slug}. However they compared PSS models to IGIMF models. The
IGIMF models restrict the number of massive stars in the synthetic
galaxies further because they impose a maximum cluster mass that depends
on the total star-formation rate. We have only imposed a cluster mass
that dependent maximum stellar mass and have shown that a CMDMSM alone
cannot be ruled out.
  
An important conclusion to draw from our models \citep[and those
  of][]{slug,danw} is that the scatter and variation of the
H$\alpha$/UV ratio is not due to the IMF filling method but it depends
more on the star-formation history of each individual galaxy. A
general trend we find is that those systems with less star-formation
in the last 10 Myr have lower ratios, while those with most of the
star formation in the last 10 Myr have higher ratios even at low mean
star-formation rates. This is because the stars responsible for
H$\alpha$ emission typically have ages of 10 Myr or less. This
indicates that any simulation that predicts the properties of a sample
of galaxies must take into account the stochastic nature of
star-formation and recognise not only that each cluster has its own
stellar content but also that each cluster has its own age independent
of the other clusters. If there are enough clusters in a galaxy this
leads to an \textit{average} stellar population. However if there are
only a few clusters the appearance of the galaxy-wide stellar
population can be very different from what might be expected for a
simple stellar population with a smooth star-formation history.

\subsection{The importance of binaries}

\begin{figure}
\includegraphics[angle=0, width=85mm]{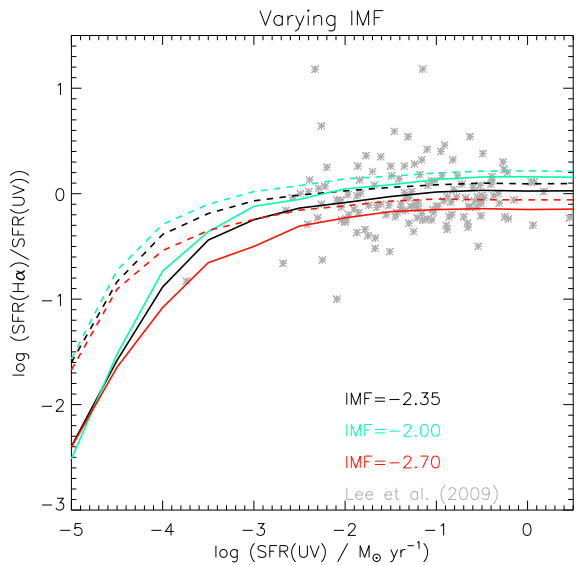}
\caption{Similar to Figure \ref{galaxymodels} but here showing the
  mean ratios for single star (solid lines) and binary star (dashed
  lines) populations calculated for PSS with different slopes for the
  SIMF. }
\label{imfplot}
\end{figure}

From our results it is possible to qualitatively demonstrate the need
to use binary star models. For individual clusters binaries seem to
have little affect. This is because of the short period of 8~Myr we
have used to match the observed clusters in this case. In our
synthetic galaxies, with 100~Myr of star-formation, we see that the
scatter of the synthetic galaxies is reduced slightly if binaries are
included and the mean SFR ratio starts to decrease at lower SFRs. This
is more clearly shown in Figure \ref{imfplot} in which we compare
populations with different IMFs and single-star to binary star
ratios. We see that for a single-star population the ratio begins to
drop between $10^{-2}$ and $10^{-3}{\rm M_{\odot} \, yr^{-1}}$ while
for binary populations this drop begins between $10^{-3}$ and
$10^{-4}{\rm M_{\odot} \, yr^{-1}}$. The binary effect makes it
difficult to distinguish between the PSS and CMDMSM IMF filling
methods at any star-formation rate.

Binary evolution affects the observed SFRs because through mass
transfer between and merging of stars it increases the number of
massive stars at the expense of lower-mass stars. We demonstrate this
in Figure \ref{imfplot}. We see here that binary populations typically
produce similar H$\alpha$/UV flux ratios to a single star population
when the IMF slope is shallower. That is until low star-formation
rates at which point, for a single cluster with a significant
population of binary stars we can also expect the apparent IMF to be
flatter. Furthermore the most massive star in a cluster might not have
been the most massive star when it formed. Therefore interacting
binary stars have a strong effect and must be included when attempts
are made to determine the IMF from observations of stellar systems.

\section{Conclusions}

We have investigated two uncertainties in population synthesis. These
are how the IMF is filled and the effects of interacting binary star
evolution.  The H$\alpha$ flux per $M_{\odot}$ observed in samples of
clusters is consistent with PSS of the SIMF for clusters around
$100M_{\odot}$ because individual low-mass clusters with one or two
massive OB and WR stars, such as the Velorum cluster or those
presented by \citet{lamb}, provide a strict test to distinguish
between PSS and CMDMSM.  For more massive clusters comparison with the
observations of \citet{calzetti} is not significant enough to rule out
CMDMSM. At these masses, around $10^3 M_{\odot}$ and above, it also
becomes more difficult to to differentiate between PSS and CMDMSM
because of the blurring effect of binary stars and in addition to the
lack of conclusive data in this mass range.

We have also considered the ratio of the H$\alpha$ to UV fluxes in
galaxies. Observationally there is a significant scatter that can be
explained by the stochastic nature of the star-formation history. We
find it difficult to differentiate between PSS and CMDMSM. This is
because we find some evidence that the leakage or loss of ionising
photons must be considered. In addition, including binary star
populations makes it difficult to distinguish between the methods for
filling the IMF. Only single-star CMDMSM populations can be ruled out
with the observations of galaxies with SFRs below $10^{-2}M_{\odot}
{\rm yr^{-1}}$.

The ratio of H$\alpha$ to UV flux for stellar populations, including
binary stars, varies less than that for populations of single
stars. Binaries can merge and mass-transfer can produce more massive
stars than were present in the initial population. Therefore the
expected star-formation rates for galaxies in which it will be
possible to detect differences between PSS and CMDMSM are much lower
than currently observed. Furthermore, because the leakage or loss of
ionising photons from young stellar populations must be considered, it
becomes even more difficult to discern the IMF-filling method from
observations of galaxies with low H$\alpha$ to UV ratios. We suggest
that it may be more fruitful to find galaxies with low overall
star-formation rates but with high H$\alpha$ to UV ratios. That is
galaxies that are rich in clusters similar to those found by
\citet{lamb}.

\section{Acknowledgements}

JJE would like to thank the anonymous referee for his very
constructive comments which have lead to a much improved paper. JJE is
supported by the Institute of Astronomy's STFC Theory Rolling
grant. JJE would also like to thank Joe Walmswell, Monica Relano, Ben
Johnson, Dan Weisz, Janice Lee, Daniella Calzetti, Sally Oey, Mark
Gieles, Mieguel Cervi\~{n}o, Michele Fumagalli, Robert da Silva and
Christopher Tout for very helpful discussions and comments on this
paper.

\bsp

%\begin{figure*}
%\includegraphics[angle=0, width=168mm]{figappx1.ps}
%\includegraphics[angle=0, width=168mm]{figappx2.ps}
%\caption{Predicted UBVRIJHK magnitude distribution for type II
%  SN progenitors at LMC and solar metallicity. Single stars in black,
%  binary stars in red.}
%\label{predictedmagnitudes3}
%\end{figure*}

\end{document}